\documentstyle[preprint,aps,prl,epsf]{revtex}

\begin{document}

\title{ Strong ferromagnetism and weak antiferromagnetism in double
  perovskites: Sr$_2$Fe{\it M}O$_6$ ({\it M}=Mo, W, and Re) }

\author{ Z. Fang$^{1}$, K. Terakura$^{2,3}$ and J. Kanamori$^{4}$ }

\address{
       $^{1}$JRCAT,
               Angstrom Technology Partnership (ATP),
               1-1-4 Higashi, Tsukuba, Ibaraki 305-0046, Japan\\
       $^{2}$JRCAT,
               National Institute for Advanced Interdisciplinary
               Research, 1-1-4 Higashi, Tsukuba, Ibaraki 305-8562,
               Japan\\
       $^{3}$Tsukuba Advanced Computing Center,
               1-1-4 Higashi, Tsukuba, Ibaraki 305-8561, Japan\\
       $^{4}$International Institute for Advanced Studies,
               9-3 Kizugawadai, Kizu-cho, Soraku-gun, Kyoto 619-0225,
Japan\\
}

%\date{\today}
\maketitle

\begin{abstract}
  Double perovskites Sr$_2$Fe$M$O$_6$ ($M$=Mo and Re) exhibit
  significant colossal magnetoresistance even at room temperature due
  to the high Curie Temperature (419K and 401K).  However, such a high
  Curie Temperature is puzzling, given the large separation between
  magnetic elements (Fe). Moreover, with $M$=W, the electronic and
  magnetic properties suddenly change to insulating and
  antiferromagnetic with the N{\'e}el temperature of only 16$\sim$37
  K.  Based on detailed electronic structure calculations, a new
  mechanism is proposed which stabilizes the strong ferromagnetic
  state for $M$=Mo and Re and is passivated for $M$=W.  
\end{abstract}

\newpage
%\maketitle

Intensive studies on the perovskite transition-metal oxides (TMO),
particularly manganites, have revealed a variety of novel phenomena
only for half a decade~\cite{ref1}.  Among those phenomena, colossal
magnetoresistance (CMR) has been attracting strong attention not only
as a challenging subject of fundamental science but also as an
important phenomenon for potential technological application.  With
regard to the latter aspect, materials with not only the half-metallic
nature but also Curie temperature ($T_c$) much higher than room
temperature is strongly desired in order to realize strong CMR effects
at room temperature.  It was demonstrated that some of the double
perovskite TMO such as Sr$_2$FeMoO$_6$ (SFMO) and Sr$_2$FeReO$_6$
(SFRO) are suitable candidates~\cite{SFMO,SFRO}.  They are half
metallic according to the band structure calculations and their
$T_c$'s are 419K and 401K.

The present work deals with two fundamental problems in these double
perovskite TMO by performing detailed electronic structure
calculations.  In both of SFMO and SFRO, the magnetic moments of Fe
are aligned ferromagnetically and the induced moments on Mo and Re are
coupled antiferromagnetically to Fe moments.  Therefore these
materials can be regarded as ferrimagnetic.  However, we regard them
ferromagnetic (FM) because Mo and Re are intrinsically non-magnetic in
the sense that their magnetic polarization cannot be sustained
spontaneously by the exchange potential on these atoms.  Actually
their negative moments (i.e., antiparallel to Fe moments) are induced
by Fe moments through the 4d(5d)-3d hybridization.  Now the first
fundamental question is why $T_c$ is so high despite the fact that Fe
atoms are very much separated with non-magnetic elements (Mo, Re)
sitting in between.  We will point out that a FM stabilization
mechanism proposed by Kanamori and Terakura~\cite{kanamori} operates
in SFMO and SFRO.  The same problem was treated recently also by Sarma
{\it et al.} for SFMO~\cite{sarma}.  The second question concerns the
striking difference of Sr$_2$FeWO$_6$ (SFWO) from SFMO and SFRO in the
electronic and magnetic properties.  SFWO is antiferromagnetic (AF)
insulator and the N{\'e}el temperature is only
16$\sim$37K~\cite{SFWO}.  Why is the W case so different from the Mo
and Re cases despite the fact that W is the 5d analogue of Mo and next
to Re in the row of the periodic table?

We will show that the stronger 2p(O)-5d(W) hybridization compared with
2p(O)-4d(Mo) hybridization is the main source of the difference
between Mo and W pushing the 5d states higher in energy and
passivating the FM stabilization mechanism in SFWO.  Deeper 5d levels
in the Re case compared with the W case cancel the effect of enhanced
p-d hybridization and restores the FM stabilization mechanism.  Note,
however, that the standard LSDA (local spin-density approximation) or
GGA (generalized gradient approximation)~\cite{GGA} cannot describe
properly the ground state of SFWO.  As Fe d states are strongly
localized in these systems, the local Coulomb repulsion $U_{\rm eff}$,
which is semi-empirically taken into account by the LDA+U
method~\cite{LDAU} in the present work, plays crucially important
roles.

We adopt the plane-wave pseudopotential method.  The 3d states of Fe,
4d states of Mo, 5d states of W and Re and 2p states of O are treated
with the ultrasoft pseudopotentials~\cite{vanderbilt} and the other
states by the optimized norm-conserving pseudopotentials~\cite{TMPP}.
The cut-off energy for describing the wave functions is 30 Ry, while
that for the augmentation charge is 200 Ry.  The crystal structures of
all the three systems are cubic with the nearest Fe-{\it M} distance
given as 3.945 {\AA}, 3.975 {\AA} and 3.945 {\AA} for $M$=Mo, W and
Re, respectively~\cite{SFRO,lattice}.  For the FM state, the number of
{\bf k}-points used in the {\bf k}-space integration is 19 in the
irreducible Brillouin zone.  For the AF state, two different
configurations, AFI and AFII~\cite{terakura}, are considered.  In the
AFI (AFII) configuration, the magnetic moments are aligned
ferromagnetically within the (001) ((111)) plane and alternate along
the [001] ([111])direction.  The number of {\bf k}-points in the AF
configuration is chosen to be equivalent to that in the FM
configuration. As for the electron-electron interaction, we adopt
first the standard GGA~\cite{GGA} and then the semi-empirical LDA+U
method~\cite{LDAU}.  The details of the implementation of the LDA+U
method in the pseudopotential scheme can be found in our previous
publication~\cite{sawada}.

Figure 1 shows a summary of the GGA calculations for three materials
Sr$_2$Fe$M$O$_6$ with $M$=Mo, W and Re in both FM and AF states.
(Note that only the results for AFII are shown here for the AF
states.) The thin solid lines denote the local density of states
(LDOS) for Fe 3d orbits and thick broken lines LDOS for 4d (Mo) or
5d (W, Re) states.  The results for $M$=Mo and Re in FM state are
basically the same as those shown in the previous
works~\cite{SFMO,SFRO}.  The oxygen p bands extends from -8~eV to
about -4~eV, the Fe majority spin t$_{\rm 2g}$ bands from about -4~eV
to -2~eV followed by the majority spin e$_{\rm g}$ bands extending up
to near the Fermi level.  In the majority spin state, the band just at
and above the Fermi level is of t$_{\rm 2g}$ character of $M$. In the
minority spin state, t$_{\rm 2g}$ states of Fe and $M$ coexist around
the Fermi level.  The formal valence of the combination of Fe$M$ is
+8, meaning that the number of d electrons per Fe$M$ is 6 for $M$=Mo
and W and 7 for $M$=Re.  In both FM and AF states, the majority spin
bands of Fe are completely filled with 5 electrons and the minority
spin bands accommodate 1 electron for $M$=Mo and W and 2 electrons for
$M$=Re.  These materials are predicted to be metallic in both FM and
AF orders with GGA calculations.  Particularly, they are half metallic
in the FM order and this half metallicity is preserved even in the
LDA+U calculation as shown later.  In Table I, the total energies for
AF states with reference to those for FM states are given for the
three materials.  The calculation for the AFI state for $M$=W suggests
that this magnetic order may have no chance of being realized in these
materials.  Therefore the AFI order will not be considered hereafter.
Clearly, the FM state is significantly more stable than the AFII state
in the GGA calculations for all the three systems.  The results in GGA
are qualitatively consistent with experimental facts for SFMO and SFRO
but inconsistent for SFWO, which is antiferromagnetic and insulating
experimentally~\cite{SFWO}.  Deferring the discussion on the stability
of ferromagnetism for SFMO and SFRO for a while, we first discuss the
problems of SFWO and how to solve them.

The LDOS for the AF state of SFWO has a very sharp peak of the Fe
t$_{\rm 2g}$ state origin just at the Fermi level.  This suggests that
the AFII state obtained in this stage may be unstable.  Although the
symmetry in the AFII state is reduced to D$_{3d}$, lift of degeneracy
in t$_{\rm 2g}$ orbits is not strong enough to split the t$_{\rm
  2g}$ band. The situation is quite similar to FeO~\cite{FeO}.  In
this case, the lattice is elongated along $<111>$ direction (even with
the GGA level treatment) and furthermore the local Coulomb repulsion
($U_{\rm eff}$) strongly enhances the orbital polarization making the
system insulating.  On the analogy of FeO, we first studied effects of
rhombohedral distortion of SFWO in GGA and found that such distortion
either elongation or contraction along $<111>$ direction simply
increases the total energy.  The cubic lattice for SFWO even in the
AFII state is actually observed experimentally.  As these analyses
suggest that there is little chance of stabilizing the AF state for
SFWO with the GGA level calculation, we applied the LDA+U method to
these materials.  Although the LDA+U method is semi-empirical, it
still provides us with some important insights into the problems.  We
set $U_{\rm eff}$ to be 4 eV and applied it only to the Fe d orbitals
for the sake of simplicity.  As was described in our previous paper,
$U_{\rm eff}$ is nonzero in a rather limited region around the Fe
nucleus and its actual value does not have definite
meaning~\cite{sawada}.

Figure 2 shows the LDA+U version of Fig.1, and Table I includes the
corresponding total energies.  The common characteristic feature in
Fig.2 is the enhancement in the exchange splitting of Fe.
Nevertheless, the electronic structure remains qualitatively the same
for SFMO and SFRO in both FM and AF states except the fact that the
weight of Mo and Re d states increased significantly around the Fermi
level.  On the other hand in SFWO, while the change in the FM state is
minor, the AF state shows a dramatic change from Fig.1 to Fig.2. The
t$_{\rm 2g}$ band of Fe splits due to orbital polarization induced by
$U_{\rm eff}$ and the occupied state in the minority spin state just
below the Fermi level is of a$_{1g}$ character.  A band gap opens up
and the AFII state becomes more stable than the FM state. The
insulating nature of the ground state of SFWO is now correctly
reproduced~\cite{bandgap}.  The fact that the N{\' e}el temperature is
only 16$\sim$37~K may suggest that the stabilization of the AFII state
in the present calculation may be overestimated.  However, the
quantitative aspect can be tuned by $U_{\rm eff}$.

Having shown the calculated results which are qualitatively consistent
with experimental facts, we start discussions on the underlying
mechanisms in relation to the two fundamental questions raised at the
beginning of the present paper.  The first one concerns the mechanism
of the strong stabilization of the FM state for SFMO and SFRO.
Recently Sarma {\it et al.} proposed an interesting explanation to the
origin of strong AF coupling between Fe and Mo in which they pointed
out strong effective exchange enhancement at Mo due to the 3d
(Fe)-4d(Mo) hybridization.  Kanamori and Terakura~\cite{kanamori}
proposed a more general idea for the mechanism where a non-magnetic
typical element located at the midpoint of neighboring high-spin 3d
elements contributes to stabilization of the FM coupling of the 3d
elements.  Figure 3(a) is a schematic illustration explaining the
mechanism.  The states of the typical element located in between the
majority and minority spin states of 3d elements are tentatively
called p states.  The key concept in this mechanism is the energy gain
contributed by the negative spin polarization of the non-magnetic
element induced by the p-d hybridization.  Such spin polarization does
not exist in the AF configuration and therefore there is no energy
gain due to the spin-state relaxation at the typical element.  In the
present problem, the 4d states of Mo and 5d states of W and Re
correspond to the p states in Fig.3(a).  The analogy is obvious in the
majority spin state in the FM order.  In the minority spin state, as
the 4d (or 5d) bands and the 3d bands are not well separated and the
Fermi level lies in the 3d bands after including the hybridization, we
need a careful analysis to distinguish the FM stabilization mechanism
discussed above and the double exchange (DE)~\cite{DE}.  Figure 3(b)
illustrates the situation corresponding to SFMO and SFRO where the $M$
t$_{\rm 2g}$ bands are slightly below the Fe ones.  We first treat the
up and down spin states separately and then consider the electron
transfer between two spin states.  The standard DE mechanism takes
account of the processes only in the first step.  As for the
hybridization between $M$ (=Mo, W, Re) bands and the majority spin Fe
bands, the total energy change caused by band shift due to the
3d-4d(5d) hybridization does not depend on the relative spin direction
between the neighboring Fe atoms up to the second order in the
hybridization matrix element $t$.  For example, the upward shift of
the up spin $M$ bands by $2t^2/{\Delta}$ in the FM state balances the
upward shift of the both spin $M$ bands by $t^2/{\Delta}$ in the AF
state where ${\Delta}$ denotes the energy separation.  Subtle features
exist in the minority spin state.  Not only the band shift but also
band broadening have to be considered.  It is obvious that the width
of the minority spin bands will be wider in the FM state than in the
AF state.  As the Fermi level lies in the minority spin bands, the
band broadening contributes to the stability of the FM state like in
the standard DE~\cite{DEcom}.  In the present problem, we have an
additional effect in the FM state coming from the electron transfer
just like in Fig. 3(a).  This electron transfer produces negative spin
polarization at $M$ atoms and contributes to further stabilization of
the FM state.  In contrast to SFMO and SFRO, the $M$ t$_{\rm 2g}$
bands in SFWO are slightly above the Fe ones.  In this case, the $M$
t$_{\rm 2g}$ bands are basically empty and the electron transfer will
not occur.  Therefore the FM-stabilization mechanism of Fig. 3(a) is
passivated for SFWO, while the DE mechanism may still be effective.
Although Table I still shows small negative spin polarization at the W
atom, this is due to the stronger 5d-3d hybridization in the minority
spin state than in the majority spin state.  We also speculate that
the main reason of considerable relative stability of the FM order for
SFWO in the GGA calculation is the rather unstable electronic
configuration in the AFII order.  Because of this, a change in the
electronic structure in the AFII state from GGA to LDA+U reduces the
energy of the AFII state dramatically.

The second question concerns the origin of the different behavior of W
from other two elements Mo and Re.  It is clear from the above
arguments that in order to answer this question, we have to clarify
the origin of the difference in the energy position of the minority
spin t$_{\rm 2g}$ bands.  We assign the p-d hybridization between
oxygen and $M$ to the main source of this difference.  As the 5d
orbital of W is more extended than the 4d orbital of Mo, the stronger
2p(O)-5d(W) hybridization pushes the 5d band, which is the p-d
antibonding state, higher in energy.  This mechanism is supported by
the fact that the p-d bonding counter part is clearly deeper for SFWO
than for SFMO (see Figs.1 and 2).  As Re has deeper 5d level than W to
accommodate one more d electron, the energy scheme for SFRO becomes
similar to that for SFMO.

In summary, we showed that the electronic structures and magnetic
ordering in the ground state of Sr$_2$Fe$M$O$_6$ ($M$=Mo, W and Re)
are properly reproduced by the LDA+U method.  A new mechanism was
proposed to explain the high Curie temperature for $M$=Mo and Re
cases.  An explanation was also given to the sudden changes in the
electronic and magnetic properties in the $M$=W case. The mechanism
proposed by us~\cite{kanamori} is very useful to predict
qualitatively the change in the magnetic states by changing
constituent elements.

We thank Tokura Group members in JRCAT for providing us with
experimental information.  Thanks are given also to Prof. N. Hamada
and Prof. D. D. Sarma for valuable discussion.  The present work was
partly supported by NEDO.

\newpage
\begin{table}[t]
\caption{For each Sr$_2$Fe{\it M}O$_6$ ({\it M}=Mo, W and Re), the first
  row shows the total energies per Fe (in meV), the second and the
  third rows list the magnetic moments (in $\mu_B$) of Fe and {\it M},
  respectively. Both the GGA and the LDA+U ($U_{\rm eff}$=4.0 eV)
  results are given. The number in bracket for {\it M}=W is the total
  energy for the AFI state.}
\begin{tabular}{c|c|c|c|c}
    &\multicolumn{2}{c|}{GGA}  &\multicolumn{2}{c}{LDA+U}  \\
\cline{2-5}
    &FM &AFII &FM &AFII \\ \hline
                 &0  &84    &0  &58   \\
{\it M}=Mo       &3.73  &3.68  &3.97  &3.96  \\
                 &$-$0.30 &0 &$-$0.39 &0 \\ \hline
                 &0  &64 (145)   &0  &-30   \\
{\it M}=W   &3.65  &3.63  &3.87  &3.68  \\
                 &$-$0.14 &0 &$-$0.22 &0 \\ \hline
                 &0 &103  &0  &52   \\
{\it M}=Re  &3.70  &3.63  &3.95  &3.91  \\
                 &$-$0.78 &0  &$-$0.86 &0\\
\end{tabular}
\end{table}

\newpage
\begin{figure}[t]
      \centering
      \leavevmode  \epsfxsize=120mm \epsfbox[50 30 550 750]{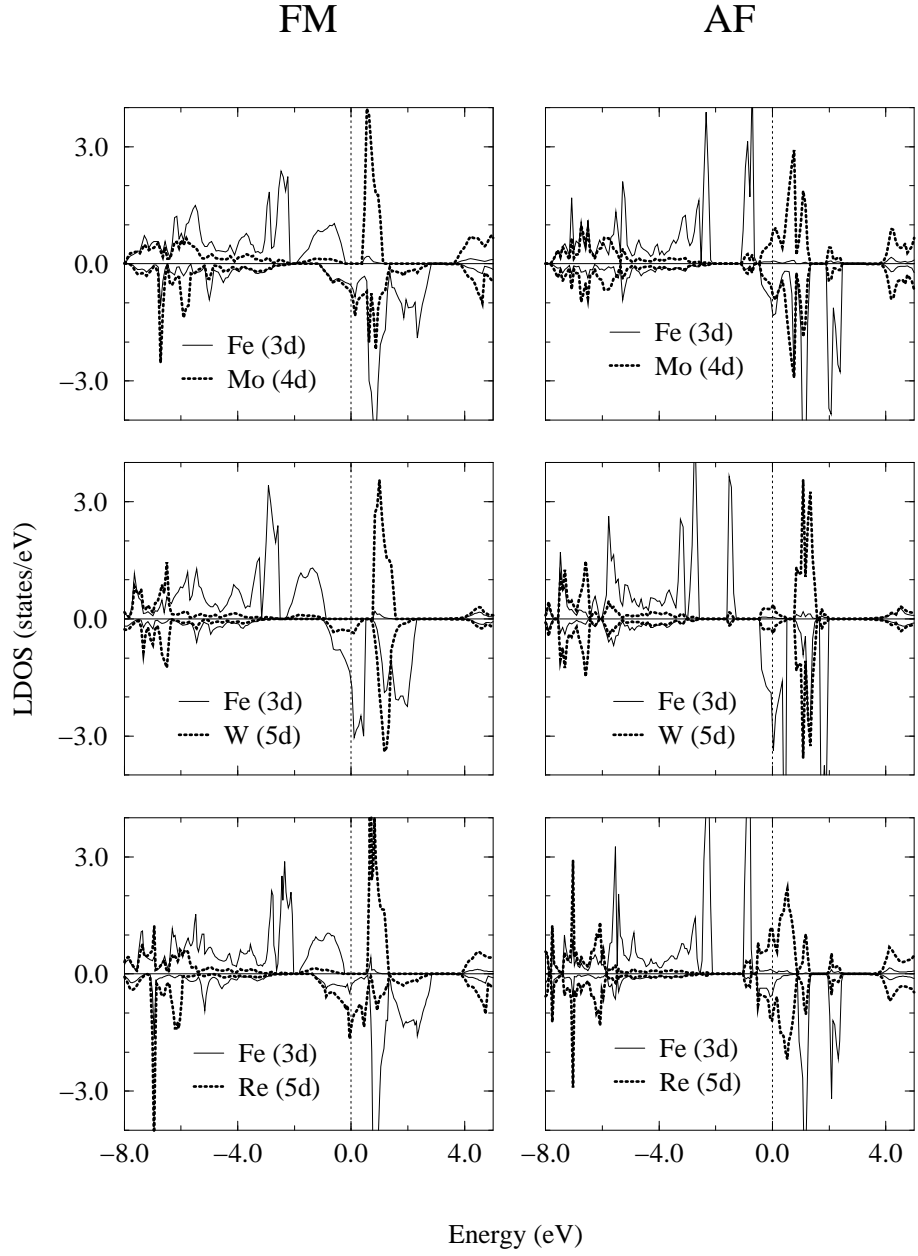}
      \caption{The calculated local density of states (LDOS)
        for Sr$_2$Fe{\it M}O$_6$ ({\it M}=Mo, W and Re) in GGA. The
        left (right) panels are for the FM (AFII) states. The energy
        zero is taken at the Fermi level.}
\end{figure}

\newpage
\begin{figure}[t]
      \centering
      \leavevmode  \epsfxsize=120mm \epsfbox[50 30 550 750]{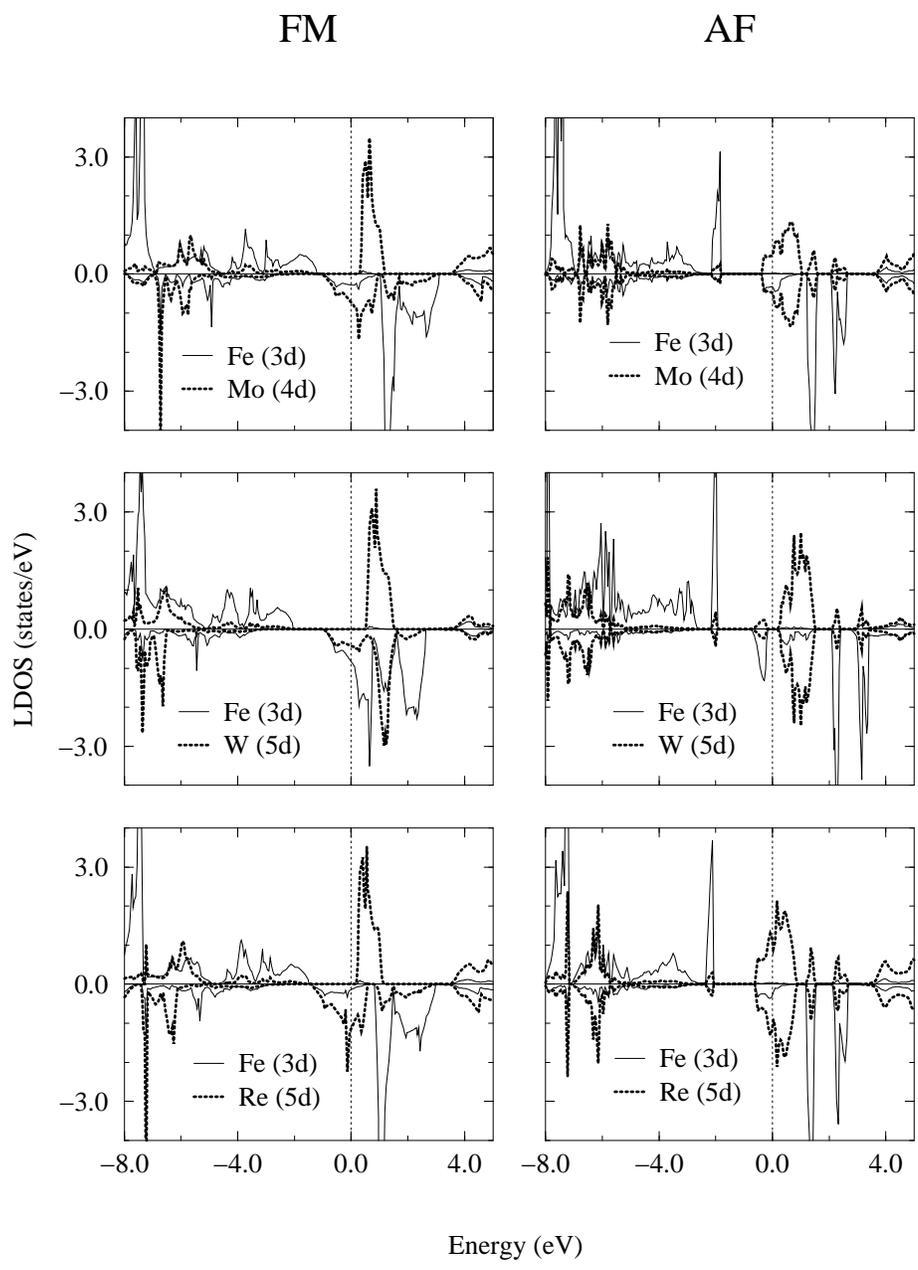}
      \caption{The LDA+U version of figure 1.}
\end{figure}

\newpage
\begin{figure}[t]
      \centering
      \leavevmode  \epsfxsize=120mm \epsfbox[50 150 550 750]{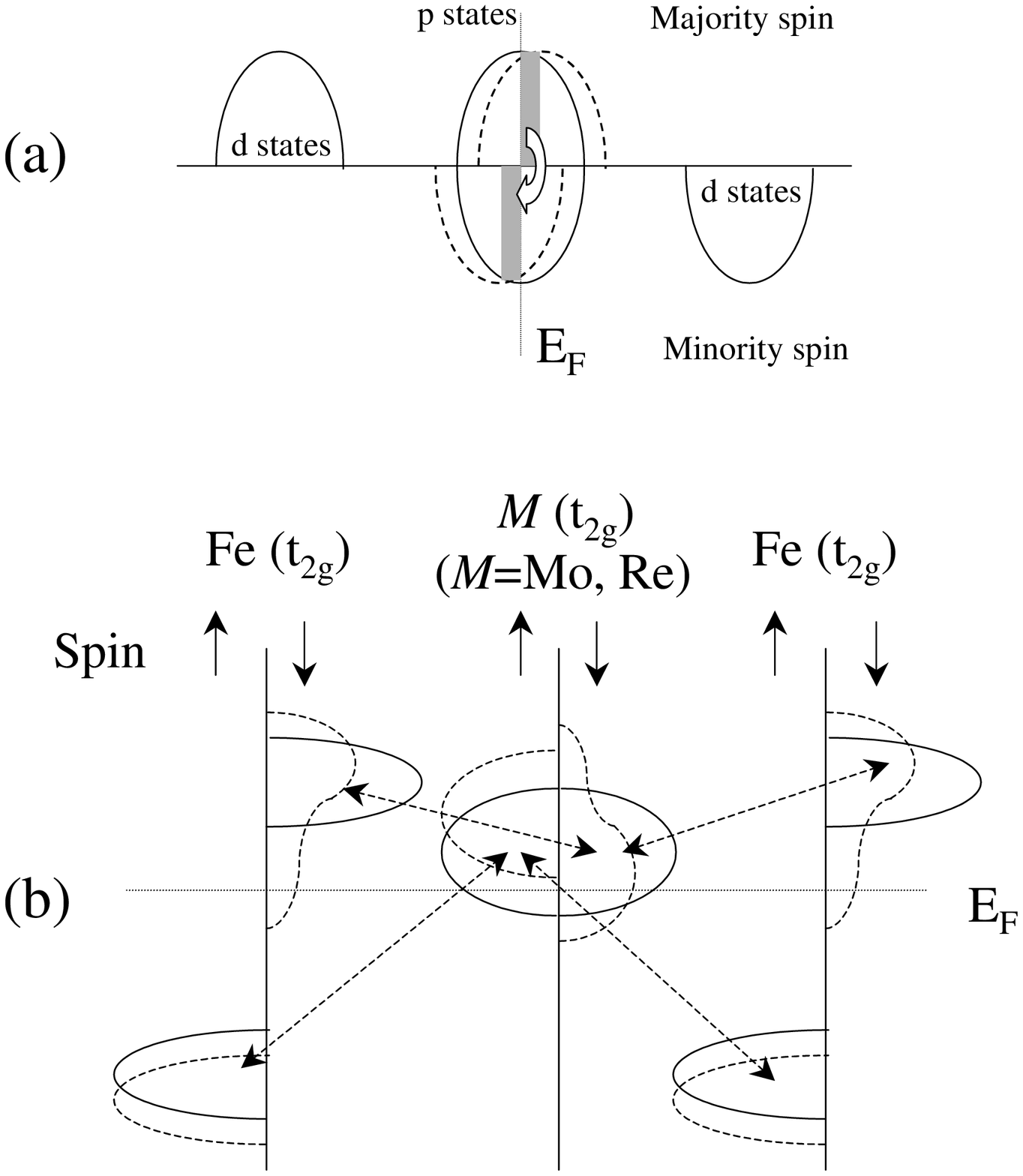}
      \caption{A schematic illustration of mechanism to stabilize
        ferromagnetic state. The panel (a) demonstrates a typical case
        for the Kanamori and Terakura mechanism, while (b) shows the
        case of Sr$_2$Fe{\it M}O$_6$ ({\it M}=Mo and Re). The
        hybridization paths are indicated by dashed lines with arrows.
        The solid (dashed) curves denote the bands without (with)
        hybridization.}
\end{figure}


\begin{thebibliography}{50}
\bibitem{DEcom} There are some qualitative difference in the DE
  mechanism between the ordinary case and the present one.  In the
  former, the transition-metal atoms interact indirectly through
  oxygens and the gain in the kinetic energy in transition-metal bands
  is the main source of the DE energy.  In the latter, $M$ is not
  simply a part of the intermediate (O-$M$-O) for the Fe-Fe coupling
  but the composite t$_{\rm 2g}$ bands formed by Fe and $M$ contribute
  to the gain in the kinetic energy.
\end{thebibliography}
\end{document}